# Catalysis distillation neural network for the few-shot open catalyst challenge

Bowen Deng

*Guangxi Key Laboratory of Petrochemical Resource Processing and Process Intensification Technology and School of Chemistry and Chemical Engineering, Guangxi University, Nanning, 530004, China*

\* To whom correspondence should be addressed. Emails: a18608202465@gmail.com (ZW)

**Abstract**

The integration of artificial intelligence and science has resulted in substantial progress in computational chemistry methods for the design and discovery of novel catalysts. Nonetheless, the challenges of electrocatalytic reactions and developing a "large-scale language model" in catalysis persist, and the recent success of ChatGPT's (Chat Generative Pre-trained Transformer) few-shot methods surpassing BERT (Bidirectional Encoder Representation from Transformers) underscores the importance of addressing limited data, expensive computations, time constraints and structure-activity relationship in research. Hence, the development of few-shot techniques for catalysis is critical and essential, regardless of present and future requirements. This paper introduces the Few-Shot Open Catalyst Challenge 2023, a competition aimed at advancing the application of machine learning technology for predicting catalytic reactions on catalytic surfaces, with a specific focus on dual-atom catalysts in hydrogen peroxide electrocatalysis. To address the challenge of limited data in catalysis, we propose a machine learning approach based on MLP-Like and a framework called Catalysis Distillation Graph Neural Network (CDGNN). Our results demonstrate that CDGNN effectively learns embeddings from catalytic structures, enabling the capture of structure-adsorption relationships. This accomplishment has resulted in the utmost advanced and efficient determination of the reaction pathway for hydrogen peroxide, surpassing the current graph neural network approach by 16.1%. Consequently, CDGNN presents a promising approach for few-shot learning in catalysis.

**Keywords:** small data; few-shot learning; catalyst challenge; catalysis distillation graph neural network; dual-atom catalyst; $H_2O_2$

**Graphical abstract**

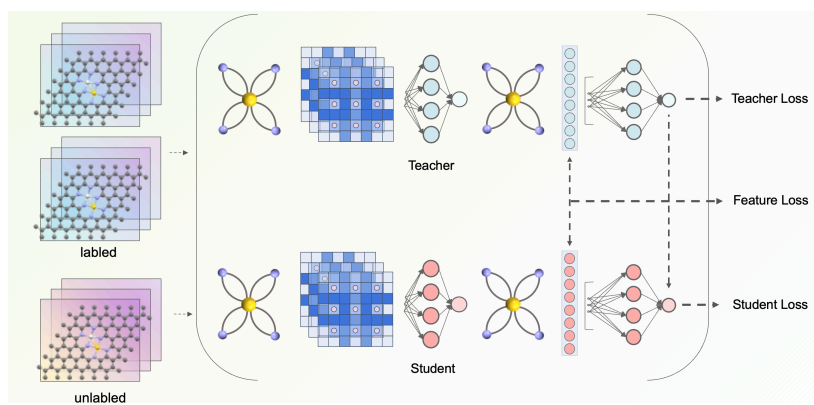

We have proposed an Few-Shot Open Challenges 2023 (FSOC23).

**Introduction**

Machine learning, an interdisciplinary field encompassing computer science, mathematics, statistics, and engineering, aims to optimize computer program performance to predict unknown sample through the utilization of data and past experiences[1]. In recent years, machine learning has found widespread applications in various domains such as imaging, text analysis, finance, healthcare, biology, and industrial processes[2,3,4,5,6]. In 2011, the Materials Genome Initiative (MGI) proposed the concept of utilizing machine learning and computational simulations to assist in material development, thereby reducing the development cycle[7]. Today, the concept of artificial intelligence (AI) for science empowers artificial intelligence in the field of natural sciences. Machine learning has emerged as a powerful tool for assisting in the design and discovery of various materials[8]. For instance, Deng et al. employed DFT+ML to propose an iterative machine learning method for predicting the adsorption free energies on different single-atom catalysts and efficiently screening ideal high-performance catalysts for $H_2O_2$ production[9]. Xie et al. proposed a graph neural network model called crystal graph convolutional neural networks, which enables the direct learning of material properties based on atomic connections within the crystal structure, thereby offering a universal and interpretable representation of crystalline materils[10]Crystal graph convolutional neural networks. It has gained increasing attention and favor from researchers, showcasing immeasurable potential. The core of machine learning-assisted catalyst design and discovery lies in the construction of well-performing machine learning models through algorithms and material data, enabling accurate predictions of target performance for unknown samples. These models can further be employed in catalyst discovery and design or to uncover hidden patterns and regularities within the data. A substantial body of research demonstrates that, compared to traditional experimental and empirical trial-and-error approaches, machine learning can rapidly extract patterns and trends from available data to guide material development, even without a comprehensive understanding of the underlying physical mechanisms.

However, significant and intricate challenges persist, particularly in the field of catalysis. Firstly, catalyst surfaces are frequently susceptible to reconstruction, leaching, doping, and defects[11], while the reaction environment can give rise to multiple potential surface terminations, resulting in highly complex data structures. Currently, no standardized coding paradigm or structure representation akin to natural language processing exists[12]. Secondly, the foremost concern lies in the substantial system sizes and the long-range electrostatic or magnetic interactions, which contribute to slower convergence. Consequently, data generation becomes highly expensive and time-consuming. Furthermore, some research data is inherently limited and most researchers tend to collect small samples under controlled experimental conditions rather than relying on large samples of unknown origin[13], which can be a significant challenge in scientific research. Thirdly, traditional machine learning algorithms usually need to map input data into a numerical vector space for processing, and their internal workings are often a complete black box, making it difficult to explain the decision-making process of the model. In contrast, Graph Neural Networks (GNNs) have stronger material structure analytic capabilities by directly expressing and processing graph structures, making them a more interpretable method[14]. Fourthly, dealing with small datasets, high or low-dimensional features, and the associated challenges of data imbalance, model overfitting, or underfitting has been a persistent issue in materials machine learning. This has long been recognized as a pain point in the field. Fifth, despite the progress made in accumulating large-scale catalyst data, the adoption of few-shot methods remains crucial for the advancement of "ChatGPT" in catalysis[15]. For instance, Jacob Devlin et al. introduced BERT (Bidirectional Encoder Representations from Transformers), a large language model based on transformer units that has been extensively applied across various tasks[16]. Due to the ongoing expansion of natural language data and the emergence of new few-shot methods, ChatGPT (Chat Generative Pre-trained Transformer) has garnered significant attention within the community, and certain industries face the risk of being displaced. Considering the aforementioned challenges, it is crucial to prioritize the exploration of few-shot techniques in small-scale catalysis data, particularly those leveraging Graph Neural Networks (GNNs).

To address these challenges, in this work, we present the Few Shot Open Challenge 2023 (FSOC23) with an example about dual-atom catalysts in hydrogen peroxide electrocatalysis and catalysis distillation graph neural network (CDGNN) as the few-shot catalysis challenge benchmark. All competition information and prize detail provided at http://172.17.0.2.

## Results

**Few-Shot Open Catalyst Challenge and Competition**. The primary objective of this challenge is to encourage greater participation from researchers in both artificial intelligence and science in the discovery of new catalysts. Due to an urgent need to design and develop green, low-cost, and high-efficiency electrocatalysts. Consequently, we propose the Few-Shot Open Catalyst Competition 2023, aimed at constructing a generalized algorithm or method to facilitate the advancement of data-driven catalysis. To accomplish this, we constructed the AFSOC23 dataset, which contains information on adsorption, structural energy, and reactions obtained from dual-atom catalysts in the oxygen reduction reaction (ORR). The dataset is constructed in three stages: (1) slab generation and relaxation, (2) adsorbate + slab generation (3) adsorbate + slab relaxation. The dataset comprises 3485 results, including isolated surfaces (also known as slabs), surfaces, adsorbate combinations (referred to as adsorbate + slabs), adsorption of broken bonds, adsorption energy, and other features. These results were subsequently divided into appropriate train, validation, and test splits. All source code and generate the adsorbate configurations will be provided in the Open Catalyst challenge repository at [http://gethub.com172.17.0.20201222](http://gethub.com172.17.0.20201222).

**Task,** The AFSOC23 dataset aims to accurately simulate atomic systems that are practically relevant to oxygen reduction reaction (ORR) applications. Presently, there is a need for the development of generalized "large language models" for catalysis and small-scale research that can benefit from few-shot methods. Therefore, the objective of the AFSOC23 dataset is to develop few-shot machine learning methods that can assist in natural science research. The competition centers on the research and exploration of key technical challenges associated with small-scale data in catalysis (materials) and aims to foster the application and development of small-scale data in catalysis (materials) in collaboration with global research talents.

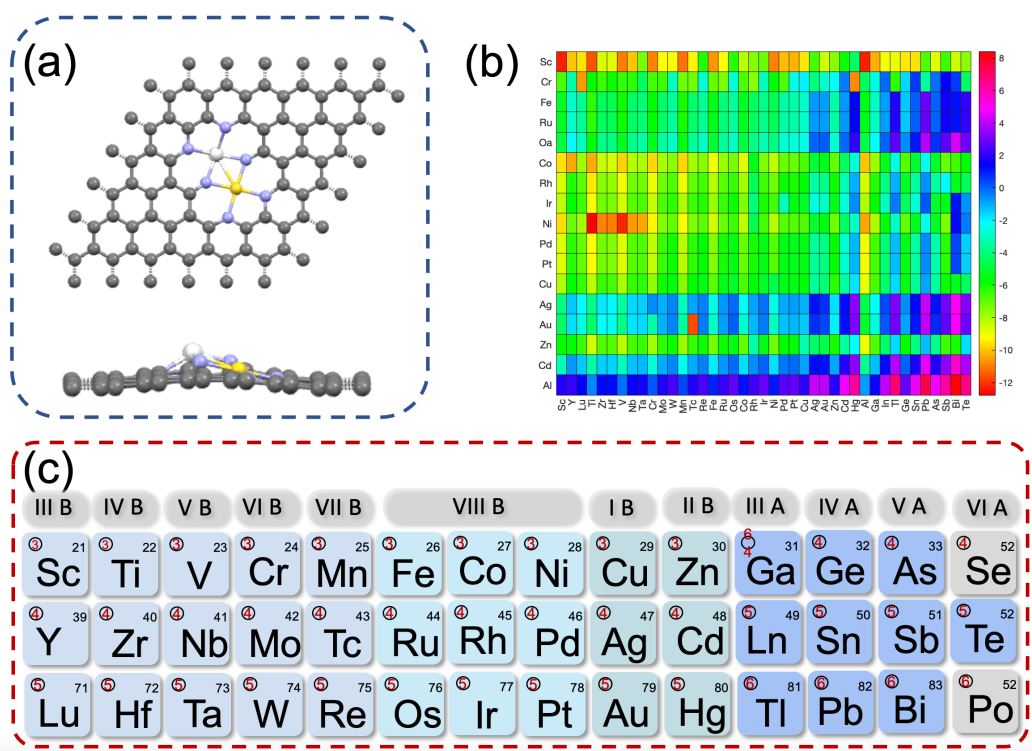

**Figure 1**. (a) Top and front view of the atomic model of the metal atom loaded on graphene. (b) The binding energy $E_b$ of the dual-atom catalysts (DACs) considered in this study. (c) The metals listed in the table are considered as potential candidates. The numbers within the circles indicate the coordination numbers of the transition metals (TMs).

**Dataset.** Recently, carbon-based single-atom catalysts (SACs) have shown significant potential in electrocatalysis[17]. The pursuit of higher electrocatalytic performance has led to the development of dual-atom catalysts (DACs) based on SACs. DACs are defined as catalysts with isolated pairs of metal atoms dispersed on a carbon skeleton, serving as active sites, and have garnered significant attention. The unique electron structure of DACs can be fine-tuned by its metal pairs, coordinate environment (CE), solvent environment, and the composition of the surrounding carbon matrix[18]. In **Figure 1a**, a porous graphene slab consisting of 62 carbon atoms is constructed, with two metal atoms (AB) placed in the center and 6 nitrogen atoms serving as the coordinating environment. The top view displays the optimized structure of the AuCo substrate graphene, while the bottom view shows the frontal perspective. The exceptional stability of DACs on the carbon surface is a fundamental characteristic of atomic catalysts. To assess the stability of DACs supported on the carbon surface, the binding energy[19] ($E_b$) of the DAC is defined as the energy difference between the bonded atomic structure and the defective structure, relative to the independent metal DAC captured at the defect position on the carbon surface (**Figure 1b**, for calculation details, refer to the Methods section). A negative value of $E_b$ indicates the stability of the DAC on the carbon surface. The more negative the value of $E_b$, the greater the stability of the DAC on the carbon surface. The horizontal and vertical axes of the graph are arranged from d-block metals to p-block metals. Most of the combinations in the upper left quadrant of the graph, which represents the d-block elements, exhibit negative $E_b$ values. Conversely, only a small proportion of combinations in the lower right quadrant, which represents a small amount of p-block elements, have $E_b$ values greater than 0. Furthermore, to evaluate the diffusion and aggregation of single atom catalyst, the cohesive energies[19] ($E_c$) of metals must be considered (for more details, refer to **Figure S3** and in Method section**)**. 41 different metals (d-zone, ds-zone, p-zone) are in our list (**Figure 1c**).

According to our calculations, the adsorption configuration of an $O_2$ molecule determines the reaction pathway leading to the formation of final products, such as $H_2O_2$ or $H_2O$. **Figure 2** illustrates the atomic representation, with light blue representing hydrogen atoms and red representing oxygen atoms. The substrate is depicted by two green atoms enclosed in a black rectangle. The left downward arrow indicates the path of ORR, in which circles 1 and 2 denote one-electron hydrogenation and circle 3 represents two-electron hydrogenation reactions. the adsorbate includes *OOH, *O, *OH and *O|||OH on two metal sites. When $O_2$ molecules adsorb on the substrate and are attacked by a proton/electron pair, three reaction pathways are formed leading to the formation of *O, *O|||OH, and *OOH, respectively. In the pathway on the left side of the diagram, *O leads to the production of $H_2O$ in both a $H^+/e^-$ pair and $2H^+/2e^-$ pairs reactions. In the pathway on the middle side of diagram, *O||OH is very unstable, leading to the breaking of the O-O bond and the formation of *O and free OH. Afterward, the *O and free OH can be attacked by a $H^+/e^-$ pair and $2H^+/2e^-$ pairs reactions, leading to formation of $H_2O$. In the pathway on the right side of diagram, *OOH is stably adsorbed on the matal site. While *OOH is adsorbed with appropriate adsorption energy, it can be attacked by $2H^+/2e^-$ pairs to generate $H_2O_2$. Thus, successful metal adsorption of ooh is essential for catalytic activity as well as hydrogen peroxide.

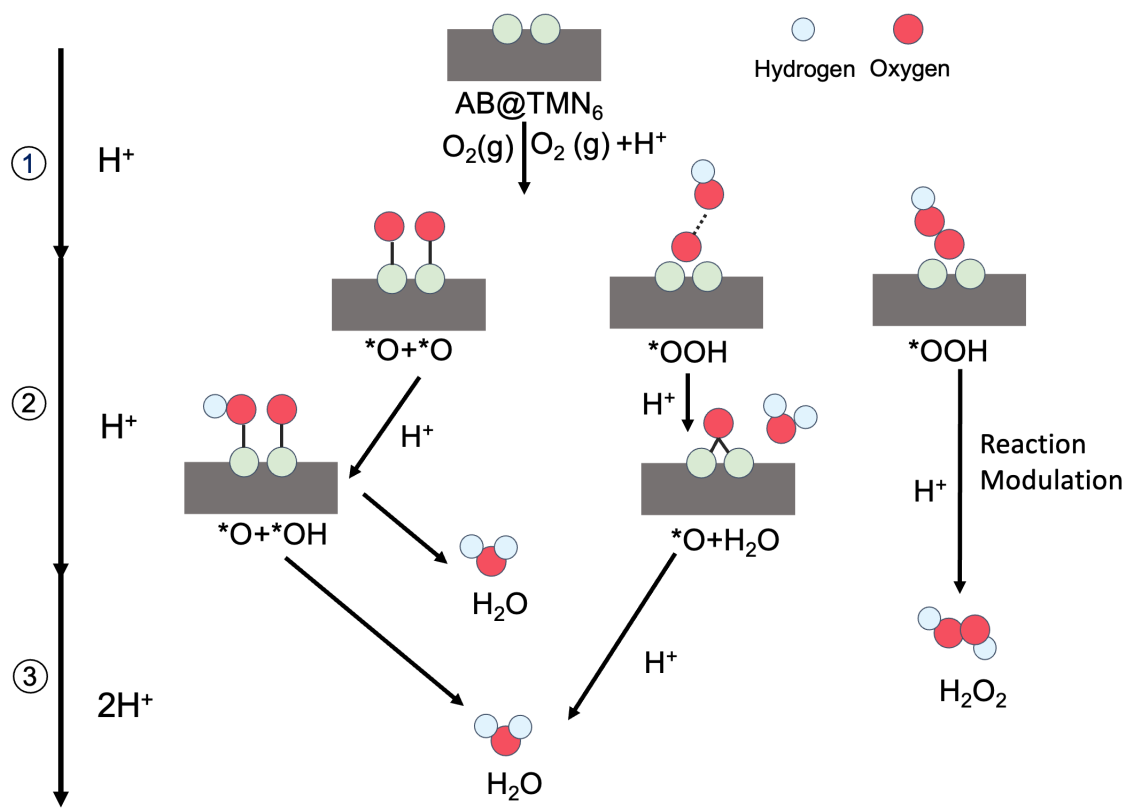

**Figure 2**. (a) Schematics of the mechanisms for oxygen reduction reaction (ORR) of $TM_a/TM_b@g\text{-}N_6$. Circle 1 and circle 2 represent a $H^+/e^-$ pair and circle 3 represent $2H^+/2e^-$ pair reaction. The left and middle parts of the diagram represent the reaction produce water and right part represent the pathway of hydrogen peroxide.

**Feature of dataset,** the adsorption free energies of *OOH intermediates on various DACs were determined using DFT calculations. We conducted a screening of the proposed DACs, considering $\Delta G_{*OOH}$ values of 4.22 ∓ 0.347 eV as the threshold for acceptable catalysts in $H_2O_2$ release (refer to **Table 1**). These DACs have the value of $\Delta G_{*OOH}$ close to 4.22 eV, including Among these DACs, $ScSc@g\text{-}N_6$, $ScV@g\text{-}N_6$, $CrLu@g\text{-}N_6$, $FeFe@g\text{-}N_6$, $FeCo@g\text{-}N_6$, $FeRh@g\text{-}N_6$, $FeIr@g\text{-}N_6$, $RuCr@g\text{-}N_6$, $OsCr@g\text{-}N_6$, $CoTi@g\text{-}N_6$, $CoMo@g\text{-}N_6$, $RhSc@g\text{-}N_6$, $RhTi@g\text{-}N_6$, $IrMo@g\text{-}N_6$, $NiY@g\text{-}N_6$, $NiTi@g\text{-}N_6$, $PdV@g\text{-}N_6$, $PtSn@g\text{-}N_6$, $PtTe@g\text{-}N_6$, $AuRh@g\text{-}N_6$, $ZnV@g\text{-}N_6$, $AuCo@g\text{-}N_6$, and $CuSn@g\text{-}N_6$ demonstrated $\Delta G_{*OOH}$ values close to 4.22 eV. It is worth noting that these DACs exhibit strong thermodynamic stability (see Supporting Information).

Table 1. The acceptable DACs with a proposed $\Delta G_{*OOH} = 4.22 \mp 0.224$ eV.

| DACs | ScSc | ScV | CrLu | FeFe | FeCo | FeRh | FeIr | RuCr |
|---|---|---|---|---|---|---|---|---|
| $G_{*OOH}$ | 4.1037 | 4.32863 | 4.37707 | 4.22305 | 4.10515 | 4.08828 | 4.0819 | 4.0248 |
| DACs | OsCr | CoTi | CoMo | RhSc | RhTi | IrMo | NiY | NiTi |
| $G_{*OOH}$ | 4.31087 | 4.23576 | 4.17282 | 4.13555 | 4.40532 | 4.10367 | 4.1216 | 4.4444 |
| DACs | PdV | PtSn | PtTe | AuRh | ZnV | AuCo | CuSn | |
| $G_{*OOH}$ | 4.2405 | 4.15202 | 4.18022 | 4.25956 | 4.12035 | 4.28524 | 4.1589 | |

Furthermore, as can be seen in **Figure 3a and 3b,** FeFe@g-$N_6$, CoTi@g-$N_6$, PdV@g-$N_6$ and AuRh@g-$N_6$ are on the weak adsorption free energy side (4.22305 eV, 4.23576 eV, 4.2405eV, 4.25956 eV> 4.22 eV) of *OOH, indicating the rate-determining step (RDS) is the formation of *OOH and the ultrahigh $U_L$ of 0.69695 V, 0.68424 V, 0.6795v and 0.65 V, respectively. Moreover, CoMo@g-$N_6$ and PtTe@g-$N_6$ are on strong adsorption free energy side (4.17282 eV, 4.18022 eV < 4.22 eV), indicating the ultrahigh $U_L$ of 0.65282 V, and 0.6602V, respectively.

Indeed, the two-electron oxygen reduction reaction (ORR) pathway has traditionally been considered a side reaction in the four-electron ORR process for fuel cell applications over the past decades. Thus, two-electron ORR selectivity on DACs can be used as a criterion for screening high-efficient catalysts for $H_2O_2$ production. **Figure 3c** illustrates the two-electron ORR activities (computed overpotential) and four-electron ORR activities plotted against *OOH adsorption free energies. The smaller overpotential observed for the two-electron ORR on DACs compared to the four-electron ORR indicates a prevailing ORR process favoring $H_2O_2$ production ( the two-electron ORR reaction needs to cross lower energy barriers than the four-electron ORR reaction, and the reaction rate of two electrons is faster, so it is more inclined to the two electron reaction path).

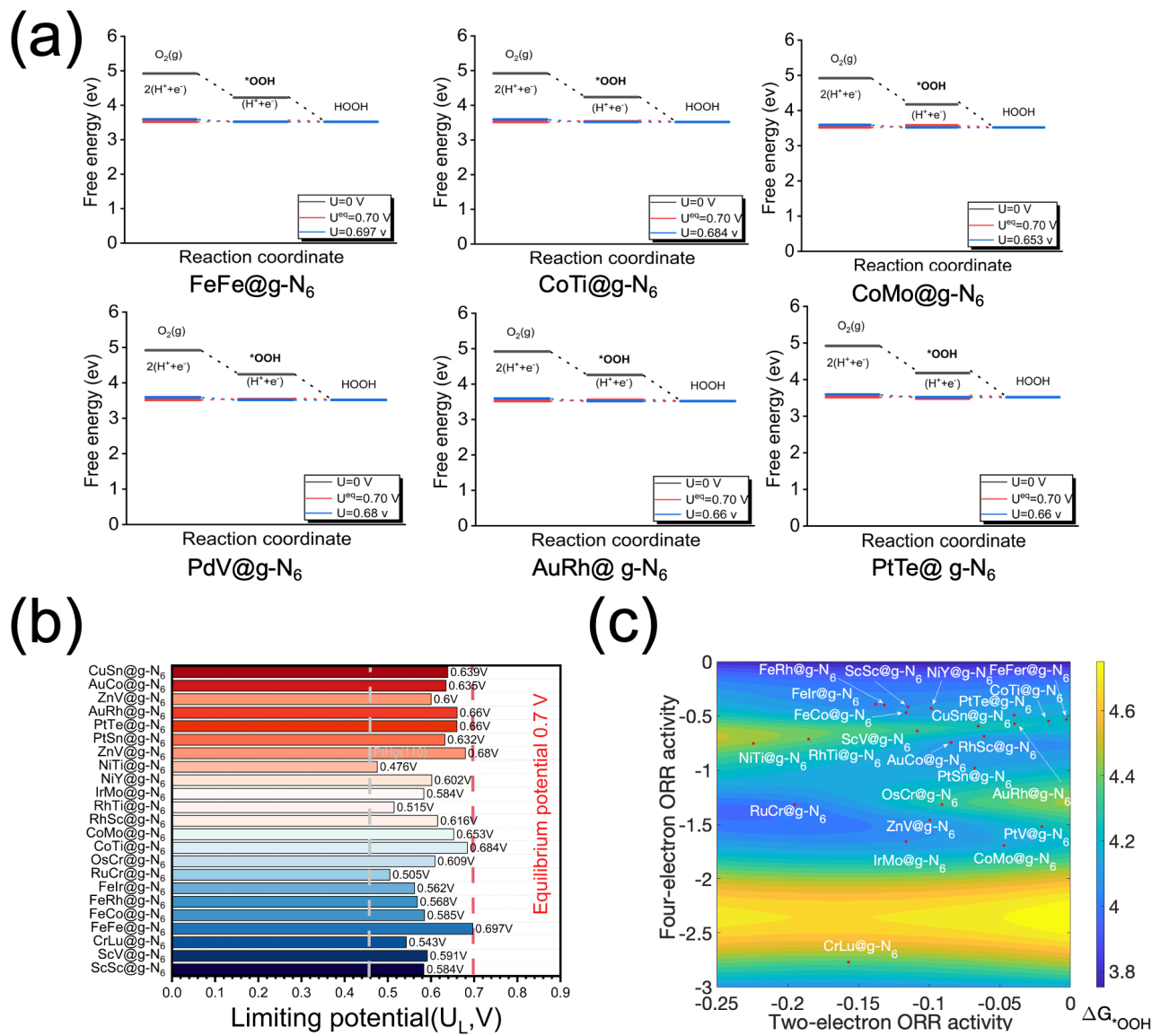

**Figure 3**. Illustration of dual-atom catalysts perfomance. **(a)** The [19] diagram of 2e-ORR on FeFe@g-$N_6$, CoTi@g-$N_6$, PdV@g-$N_6$, AuRh@g-$N_6$, CoMo@g-$N_6$ and PtTe@g-$N_6$. (b) Computed $U_L$ values of 23 DACs in comparison with the PtHg4 benchmark (UL = 0.46 V). (c) Theoretical volcano for the two-electron and four-electron oxygen reduction reaction processes as a function of *OOH adsorption free energy.

**Benchmark**. Few-shot catalysis network. In this paper, we propose the catalysis distillation learning network, which combines a graph neural network (GNN) and an architecture based exclusively on Multi-Layer Perceptron (MLP) for a few-shot catalysis baseline. The CDGNN (catalysis distillation learning catalysis graph neural network) utilizes unlabeled and a few-shot labeled data to address the challenges posed by an inadequate number of labeled samples and the significant dearth of negative samples, thereby optimizing the potential of the available data. CDGNN is composed of two neural network models which are trained in parallel, as show in **Figure 4a**: a teacher network and a student network. These two models are trained in parallel. The teacher model generates pseudo labels on the unlabelled data which are then used to train the student network. The teacher model is trained with two objectives in our case: labeled data performance and a feedback signal from the student model based on its performance on the labelled dataset. This feedback signal provides a guide for the teacher model in the case when the unlabeled samples are unlike the labeled data. The student model is trained only on unlabeled data with hard pseudo-labels provided by the teacher model. This leverages unlabeled data to improve further than supervised learning and smooth biases that may be found in the labeled data, such as through imbalances as with formation OOH classification. The CDGNN model, which serves as a wrapper framework for classification and can be combined with any structure property prediction model. Here apply the CDGNN model to forecast the successful adsorption of *OOH at the adsorption site of the dual-catalyst as a example. the successful adsorption of *OOH in catalytic generation and 2e reaction of hydrogen peroxide is important for the efficient conversion of water and oxygen to $H_2O_2$.

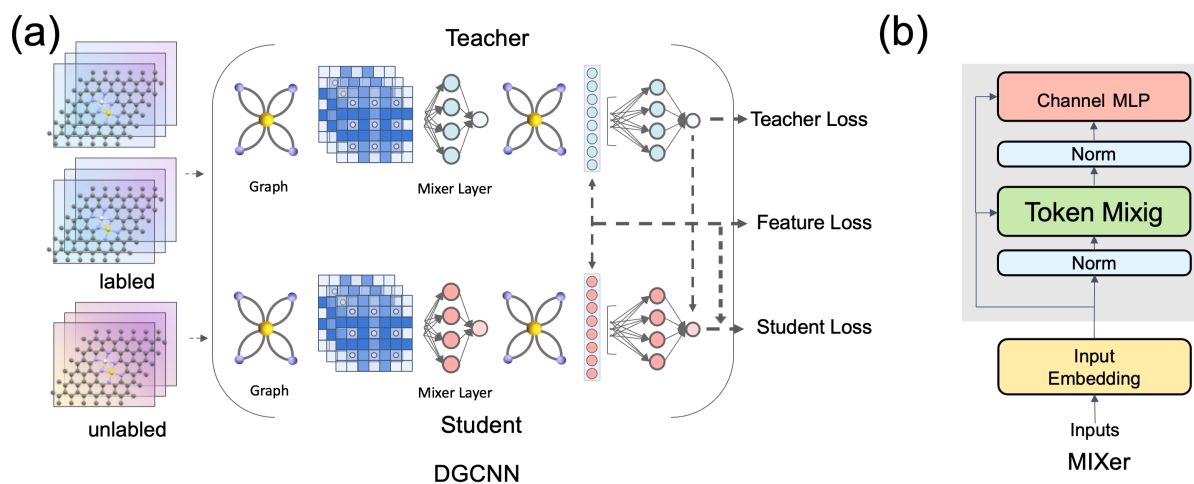

**Figure 4**. Illustration of catalysis distillation learning catalysis graph neural network. **(a)** converting catalysis structures in their unit cell into catalyst graphs by encoding atoms as nodes and bonds as the edges between them. The framework of CDGNN network contain one teacher submodule and one student submodule. (b) Construction of the MLP-Like unit with one token-mixing MLP and one channel-mixing MLP, each consisting of two fully-connected layers and a GELU nonlinearity.

In **Figure 4a**, the CDGNN model operates by transforming the unit cell of catalyst structures into undirected graphs, wherein atoms are depicted as nodes, and the edges represent the bonds between them. The subgraph of each catalyst is represented by a feature vector that is generated by a combination of catalytic and geometric feature of its elemental identity. These atomic features are encoded as one-hot encoding, and bond interactions are defined by their neighbours within a radius of 8 Å. The model can learn the intrinsic structural characteristics. After the delineation of subgraphs, each node and edge attribute is expanded as a vector representation of the user-defined chemical and geometric features. To systematically capture the chemical and geometric environment surrounding each node, the node feature vector for each node in a subgraph is iteratively updated through multiple rounds of Mixerblock, as shown in **Figure 4b**, based on its neighboring environment. Mixerblock is the main building block of CDGNN, which contains two subnetworks, a token mixer and a channel mixer. The token mixer aggregates the neighboring atoms' features, while the channel mixer aggregates the atom's features. Residue connection and layer normalization is adopted for both the attention and the MLP. The outputs of the two subnetworks are then multiplied element-wise and summed up to obtain the final representation of the atom. The training process, a batch of labeled and a batch of unlabeled data are initially sampled. The teacher's loss is then computed on the labeled batch by dkd (decoupled knowledge) loss and dynamic loss. pseudo-labels are generated by unlabled data from teacher's model. The unlabeled loss of student will be computed by the pseudo-labels. The student will update with pseudo-labels's loss and the mse (mean square error) loss between the last layer of teacher model and student model. The teacher model is then updated by combining the feedback signal with its loss on the labeled data. This method enables the student model to learn the true labels of a large set of unlabeled data, without being influenced by biases from the labeled dataset. Overall, this approach represents a powerful technique for effectively training machine learning models.

Loss functions of our student and teacher network include:

Student:

$$\mathcal{L}_u^s = CE\big(T(x_u), S(x_u)\big) + MSE\big(fea(x_u), fea(x_u)\big) \qquad (1)$$

Where $\mathcal{L}_u^s$ represents the cross-entropy loss CE and feature loss on a batch of unlabeled dataset for $[S(x_u), fea(x_u)]$ the student S network with respect to the labels produced by the teacher $[T(x_u), fea(x_u)]$. This is the student's only loss function.

Teacher:

$$TCKD = KL(b^\tau || b^s) \qquad (2)$$

$$NCKD = KL(P^P || P^s) \qquad (3)$$

$$DKD = \alpha TCKD + \beta NCKD \qquad (4)$$

Where TCKD (Target Class Knowledge Distillation) is respect to the binary probability distribution involving only target categories versus non target categories, and NCKD (Non-target Class Knowledge Distillation) represents the KL Divergence and Weights of Non-objective Probability Distributions. DKD represents the sum of $\alpha TCKD$ and $\beta NCKD$, the $\alpha$ and the $\beta$ are hyperparameters.

**Table 2**. Comparison of classification performance for *OOH.

| Model | Precision | Accuracy | Loss |
|---|---|---|---|
| **CDGNN** | 85% | 85% | **0.612** |
| **XGBoost** | **78.57%** | 79.28% | **NAN** |
| **CGMLP** | 71.9% | 71.9% | **0.729** |
| CGCNN | 68.9% | 68.9% | 2.4 |
| SVC | 66.67% | 66.67% | NAN |
| MLP | 61.24% | 61.24% | 0.6753 |

**\*OOH classification performance.** We use 612 labeled and 90 unlabeled as training data for our model CDGNN and same 612 labeled data to train CGCNN for comparison. **Table1** shows the classification performance of two models on our test set of catalysts. Our model CDGNN achieves 85.0% precision compared to the CGCNN model precision of 68.9%. with a

significant absolute 16.1% improvement by using our method. At the same time, this model achieved a loss of 0.4739, with an absolute 1.1 reduce over CGCNN model. Since this is a binary task, the reduction in loss is not reflected in the accuracy. However, the reduction in loss means that the model has better discrete performance for the dataset and learns deep semantic information about the dataset. Furthermore, in order to establish baselines for the Few-Shot Open Challenge 2023, we present results obtained using three approaches. Thus we labeled 90 unlabeled data instances and used them for training traditional machine learning models. the current machine learning easily shows excellent performance in the training set under multiple epoch rounds therefore the performance criteria were performed on the test set (Table 2). The models trained in the test set, XGBoost performed the best with precision values of 78.5 and accuracy values of 79.28%, during the validation period. SVC and MLP have very low accuracy in the test set with 66.7% and 73.3% respectively. And the loss is 0.7 and 2.47 respectively. the difference is that MLP and SVC have high accuracy in the test set with 80.7% and 66.67% respectively.

**Discussion**

In summary, machine learning based catalytic property prediction faces the big challenge of lack of sufficient annotated property data, the issue of missing negative samples (non-stable materials), expense computing resource, time-consuming and the interpretability of the structure-activity relationship are required by exhaustive first principles analysis, which is indispensible for building screening models for new catalysts discovery. To address above issues, we propose a MLP-Like based model, named GMLP, for aggregating atomic feature and their neighbours. We also introduce a knowledge distillation framework named catalysis distillation graph neural network (CDGNN) for catalysis property prediction as examples, achieves state of the art. Our extensive experiments show that our CDGNN are able to significantly improve the prediction performance compared to previous methods in catalyst adsorption. Furthermore, in order to attract more researchers in artificial intelligence and catalysis to pay attention to the few-shot challenge in the field of catalysis and permit the more extension of machine learning-based approaches to catalytic, we propose Few-Shot catalysis challenge, more information and prize detail about this challenge can be available at http://172.17.2.2221. Our work sheds light on artificial intelligence in few-shot science. In the future, artificial intelligence is able to expand to a wider range of sciences of data.

**Method**

**DFT method**, All the calculations were carried out by using spin-polarized density functional theory (DFT) method and performed by the Vienna ab initio Simulation Package (VASP) at the generalized gradient approximation (GGA)/Perdew-Burke-Ernzerhof (PBE) level. The projector augmented wave (PAW) potentials were applied. The plane wave cutoff was set to 500 eV. The electron convergence energy was set to $10^{-6}$ eV and all the atomic positions were allowed to relax until the forces were less than 0.02 eV/Å. The Brillouin zones were sampled by a Monkhorst-Pack k-point mesh with a 2×2×1 for the $CoN_4$ configurations. To avoid the interactions between two adjacent periodic images of geometries, the vacuum was set to 15 Å. The DFT-D3 scheme was adopted to correct the van der Waals interaction.

The Formation energies (in eV/atom) of dual-atom catalysts were determined based on the expression in Eq1. where $E_{surface}$, $E_{(@N6-C)}$, $E_{(Mi-bulk)}$, and $N_i$ represent the energy of the catalyst surface, the energy of the surface without dual-metal sites, the energy of the unit bulk crystal of the corresponding transition metal, and the number of metal atoms in the bulk, respectively.

$$E_f = E_{surface} - E_{(@N_6-C)-} \sum_{i=1,2} E_{(Mi-bulk)}/N_i \tag{5}$$

The binding energy of dual-atom catalysts were calculated by Eq5, where $E_{DACs}$, $E_{g\text{-}CN}$, and $E_M$ are the energies of DACs, pure g-CN and metal atoms;

$$E_b = (E_{DACs} - E_{g-CN} - nE_M)/n \tag{6}$$

The cohesive energy of dual-atom catalysts were calculated by Eq6, where $E_{M\text{-}bulk}$ is the energy of metal crystal and n is the number of metal atoms in the crystal.

$$E_c = (E_{M-bulk} - nE_M)/n \tag{7}$$

**During the training process**, the data is randomly split into four subsets: training (labeled subset and unlabeled subset (15%) ), validation, and test sets. The test set is reserved for final evaluation, while the training sets are used for model training and hyperparameter tuning. We analyzed data drift and split stochasticity by considering various random seeds in the data splitting process, but we did not observe any discernible impact on the performance of the model. We implement our network with pytorch framework. The models are trained for 100 epochs using Adam with warm-up for training. The MLP-Mixer unit is set 3. In the machine learning algorithm training process, labels are added to the unlabeled data of the gnn algorithm training set.

Unlike CDGNN, other methods differ in their approach to handling data. They assign labels to a portion (15%) of the unlabeled data for training the model, resulting in the data is randomly split into tree subsets: training labeled subset (add unlabeled subset (15%) ), validation, and test sets. We use scikit-learn and the SVC and XGBoost algorithm implementations provided by the XGBoost package. All machine learning algorithms consider 17 different physical and chemical properties, including ((1) The electron numbers of d and p orbital (Edp); (2) Oxide formation enthalpy (Hfox), (3) Pauli electronegativity (Nm); (4) Electron affinity (Am); (5) The first ionization energy (Im); (6) Atomic number (An); (7) Atomic radius(R); (8) Valence electron (Ve); (9) valence electrons in the occupied d orbitals of the (Nd); (10) valence electrons in the occupied p orbitals of the (*Np*) are selected from database. All experiments are running on a workstation with a dual-Tesla P100 GPU, and dual-Intel(R) Xeon(R) CPU(E5-2630 v4 @ 2.40GHz).

**Evaluation criteria.** We evaluate the CDGNN based on their true positive rate on each model's respective test set. We use a prediction score boundary of 0.5 to determine a positive or negative sample classification. The terms TP (True Positive), TN (True Negative), FP (False Positive), and FN (False Negative) are used to describe the performance of a classification model. TP refers to the number of correct positive predictions, TN refers to the number of correct negative predictions, FP refers to the number of incorrect positive predictions (also known as "false positives"), and FN refers to the number of incorrect negative predictions (also known as "false negatives"). These metrics are essential for evaluating the accuracy of the model, assessing its strengths and weaknesses, and improving its performance.This classification performance can be expressed as 4. We evaluate the DLCGNN on three metrics with variable formation energy thresholds: accuracy, precision, and F1 Score. We again use a prediction score boundary of 0.5 to determine a positive or negative sample classification. The accuracy metric is shown as 5. The precision and recall metrics can be expressed as 6.

$$ACC(T) = \frac{TP + TN}{(TP + FN) + (TN + FP)} \quad (9)$$

$$PR(T) = \frac{TP}{TP + FP} \quad (10)$$

$$F - score = \frac{2 Precision * Recall}{Precision + Recall} \quad (11)$$

**Leaderboard**. In order to maintain consistent and fair evaluation, a public benchmark is provided on the Few-shot Open Catalyst Project webpage (http://github.com172.17.0.1). Participants can upload their results on the test datasets for evaluation. However, the ground truth test data is not publicly released to prevent potential overfitting. Evaluation on the test set can only be conducted through the benchmark platform. Ablation studies are also encouraged to further analyze and understand the performance of different approaches.

**Data availability**

The optimized high coverage configurations in form of raw file and process datasets for the Few-Shot Open Challenge are available at: https://172.17.0.2

**Code availability**

The CDGNN code as implemented using PyTorch 2.0 is available from: https://github.com made available under MIT license.